\documentclass[11pt, oneside]{article}   	
\usepackage{geometry}                		
\usepackage{graphicx}				
\usepackage{amssymb}
\usepackage{amsmath}
\usepackage{amssymb}
\usepackage{mathtools}
\usepackage[superscript]{cite}

\title{Point Spread Function for Ground Layer Adaptive Optics}
\author{Donald Gavel}
\date{}							

\begin{document}
\maketitle

\begin{abstract}
A ground-layer adaptive optics system (GLAO) uses a single adaptive mirror to partially correct the wavefront for atmospheric and telescope aberrations over a wide field of view. Instead of reaching diffraction limit on a narrow field, the idea is to provide partial improvement of the PSF over a wide field appropriate for multi-object spectrographs or wide field imagers. We derive an analytic formulation for the GLAO corrected PSF and then apply that in developing a methodology for calculating sensitivity improvement for astronomical instruments.
\end{abstract}

\section{Introduction}

The sensitivity and signal-to-noise in astronomical instrumentation is critically dependent on the point-spread function (PSF) of the optics. At a ground-based observatory, the atmosphere is the main determinant of the PSF, so astronomers hope for good seeing nights where the PSF of the atmosphere has a narrower full-width-half-max. Adaptive optics (AO) systems are designed to correct for the atmospheric aberrations, essentially to effect an improved seeing. However, AO does not simply modify seeing, i.e. narrow the atmospheric PSF, but instead produces a more complicated point-spread distribution of imaged light that must be evaluated in the context of the AO correction limits and the seeing.

High-Strehl AO systems produce a much different PSF than the free atmosphere. These have a core-halo structure, with the core being like the diffraction-limited Airy function and the halo being somewhat of a Lorentzian shape with extent roughly the seeing disk. To achieve high Strehl requires the wavefront be corrected to  on the order of $\lambda / 4$ mean square average over the aperture. This can only be accomplished (with a single correction element) over a narrow field of view since the wavefront difference over the field quickly exceeds this limit due to the upper-altitude components of aberration.

In the case of GLAO, wavefront correction is designed to preferentially remove the lower altitude (ground layer) contributions to the wavefront. This results in partial AO correction usable over a wide field of view. At most telescope sites, the ground layer is the dominant contributor to total wavefront aberration, so GLAO will significantly reduce the total aberration. The remaining aberrations, from upper layers, are still not so small that the residual wavefront is phased to form an Airy PSF, yet there can be significant improvement in the compactness of the PSF and the more compact PSF results in improved instrument sensitivity. GLAO works by averaging wavefronts from several directions.  The wavefront contributions from upper layers tend to average to zero (variance reducing as $\sqrt{N}$ where $N$ is the number of guidestars), while the common ground layer component is enhanced, with a similar reduction in variance.

In this paper we develop the analytic formulas for the time-average (long-exposure) GLAO PSF in terms of GLAO parameters and seeing conditions. The formulas depend on seeing profile, positions of guide stars, and conjugate height of the deformable mirror. Additional terms are derived that account for wavefront control imperfections such as finite sampling of the wavefront across the deformable mirror and temporal response of the AO controller.

We then discuss how these results can be used to calculate sensitivity improvements for astronomical instruments.

\section{Mathematical preliminaries}

\subsection{Point-spread of static aberration}

In a static optical system, the point-spread function (PSF) from the object to the image plane is determined by the optical aberrations of the intervening optics. Aberrations are expressed in radians phase variation, and are given at the pupil plane as $\psi({\bf x})$ (where ${\bf x}$ are the coordinates on the pupil plane). The point-spread function in the far-field resulting from static aberrations in the near field is given by the Fraunhofer approximation
\begin{equation}
\label{eqn:PSFinst}
PSF_{inst}(\theta) = \left|{\cal F} \left\{ P(x) e^{i \psi(x)} \right\}\right|^2 = \left| \int{P(x) e^{i \psi({\bf x})}\; e^{-i 2 \pi {\bf x} \cdot {\bf \theta} / \lambda}\; d^2 x} \right|^2
\end{equation}
where $\theta$ is position in the focal plane and $P(x)$ is the pupil function
\begin{equation}
P(x) = \begin{cases}
                      1 \quad \text {$x$ inside the telescope aperture} \\
                      0 \quad \text {$x$ outside the telescope aperture}
           \end{cases}
\end{equation}

Note that the Fraunhofer expression involves a Fourier transform, mapping spatial variable $x$ to a spatial frequency variable $\theta/\lambda$. At this point it is useful to define the modulation transfer function, MTF, since it will be useful in the subsequent discussion. The MTF is the Fourier transform of the PSF:
\begin{align}
MTF(x) &= \int PSF(\theta) e^{i2\pi {\bf x} \cdot \theta / \lambda} \; d^2 \theta \\
PSF(\theta) &= \frac{1}{\lambda^2} \int MTF(x) e^{-i2\pi {\bf x} \cdot \theta / \lambda} \; d^2 x
\end{align}
Using Parseval's theorem and the definition of the PSF given in equation (\ref{eqn:PSFinst}), the MTF of a statically aberrated wavefront is
\begin{equation}
MTF(x) = \int P(x') P(x'+x) e^{i (\psi(x') - \psi(x'+x))} d^2 x'
\end{equation}
and in the special case where the aberration is zero, the MTF is the convolution of the pupil with itself:
\begin{equation}
\label{eqn:MTF0}
\tau_0(x) \overset{\Delta}{=}  MTF_0(x) = \int P(x') P(x'+x) d^2 x'
\end{equation}

\subsection{Point-spread of atmospheric aberration}

For a ground based telescope, random fluctuations in the index of refraction of the atmosphere above it contribute to aberration. The integrated wavefront is then
\begin{equation}
\label{eqn:on_axis_wf}
\psi(x) = (2 \pi / \lambda) \int{ n(x;z) dz }
\end{equation}
where $\lambda$ is the wavelength of the light, $n({\bf x};z)$ is the index fluctuation at horizontal position ${\bf x}$ on the beam at altitude $z$, and the integral is vertical through the atmosphere starting at telescope aperture. To reduce extra notation in the following we will make use of the wavenumber, $k = 2 \pi / \lambda$

The statistical average MTF is given by\cite{Fried1966}
\begin{equation}
MTF_{ave}(u) = \tau_0(u) e^{-\frac{1}{2}{\cal D}_{\psi}(u)}
\end{equation}
where the first factor $\tau_0$ is the MTF of the telescope aperture alone (\ref{eqn:MTF0}), and the second factor is a result of the atmospheric turbulence, which depends on the structure function
\begin{align}
{\cal D}_{\psi}(u) &= \left< ( \psi(x) - \psi(x+u))^2 \right>_{{\bf x}, t}
\end{align}

The structure function is a second order statistic that depends on the distance between two points across the beam in the horizontal plane. The wavefront is assumed statistically stationary, so that it doesn't matter where or when ($x$, $t$) the difference is taken. The structure function and the auto-correlation function, ${\cal C}_{\psi}(u) = \left< \psi(x)\psi(x+u) \right>_x$ are related through
\begin{equation}
\label{eqn:DCrelation}
{\cal D}_{\psi}(u)  = 2\; {\cal C}_{\psi}(0) - 2\; {\cal C}_{\psi}(u) 
\end{equation}

For fully-developed atmospheric turbulence, the structure function has the particular power-law form [Kolmogorov] given by:
\begin{equation}
\label{eqn:SFpower}
{\cal D}_{\psi}(u) = 2.914\; \left | u \right | ^{5/3}  k^2 \int C_N^2(z) dz 
\end{equation}
where $C_N^2(z)$
is the structure ``constant,'' which quantifies the strength of the turbulence at each layer of altitude $z$.

We will also find it useful to describe the structure function of refractive index at a single layer:
\begin{equation}
\label{eqn:SFindex}
{\cal D}_n(u;z) = \left< \left( n(x;z) - n(x+u) \right)^2 \right> = 2.914\; C_N^2(z) \left|u\right|^{5/3}
\end{equation}

\section{GLAO wavefront correction}

Turbulence tends to be stronger near the ground, therefore if the ground component of turbulence is removed with adaptive optics then one expects a high payoff in wavefront improvement.
We now investigate the PSF that results after GLAO correction. We start with the idealized case where wavefront measurements from multiple guidestars from several locations on the field are combined and applied to a deformable mirror perfectly and without delay. Hence the PSF in this model is determined solely by the seeing conditions and the locations of the guidestars.  In subsequent sections, we will evaluate additional effects on the PSF that are caused by the imperfect spatial and temporal frequency response of the AO system.

\subsection{Combining multiple guidestars}

The GLAO approach is to collect measurements of the wavefronts from guide stars in multiple directions, $\alpha_j$ $j = 1,2,\ldots N$ over the field of view, then apply some sort of average correction to the deformable mirror.  The thinking is that the averaging of wavefronts coming from different directions will tend to cancel out contributions from upper altitude turbulence, where the beams do not overlap and so wavefront contributions add incoherently, but near the ground, where the beams do overlap the wavefront contributions add coherently. If the correction is only for the ground layer, it is not specific to any field direction (to the extent beams overlap in the ground layer).  Since a large fraction of the total atmospheric turbulence is located near the ground this correction can be very significant, albeit have less fidelity than would be if tuned to a specific field position.

The corrected wavefront is:
\begin{equation}
\label{eqn:residual_wf}
\psi_r(x; \alpha) = \psi(x;\alpha) - \sum_{j=1}^{N} w_j \psi(x;\alpha_j)
\end{equation}
where $x$ is the position on the pupil, $\alpha$ is the position in the field, the subscript $r$ denotes residual and subscript $j$ indexes the guide stars, located at field positions $\{\alpha_1, \alpha_2, \ldots, \alpha_j \ldots \alpha_N\}$. We have slightly generalized the algorithm by allowing for weights, $w_j$, in the averaging of the measurements, under the possibility that an asymmetric weighting might be more optimal, a topic to be explored later. We assume $\sum_j w_j = 1$. For simplicity at this starting point, we'll assume that each of the guide star beams are columnated, akin to using natural guide stars at infinity:
\begin{equation}
\label{eqn:off_axis_wfs}
\psi(x;\alpha_j) = k \int n(x + \alpha_j z) dz
\end{equation}
where $k = 2 \pi / \lambda$.  To account for when the beacons are laser guide stars at finite altitude $z_{B}$, we need to account for the cone-beam projection according to
\begin{equation}
\psi(x;\alpha_j) = k \int_0^{z_B} n \left(x  \left [1 - \frac{z}{z_B}\right] + \alpha_j z \right ) dz
\end{equation}
This generalization is straightforward but details will be left to a later work.

To calculate the PSF, we substitute (\ref{eqn:on_axis_wf}) and (\ref{eqn:off_axis_wfs}) into (\ref{eqn:residual_wf}) and carry through with deriving the structure function for $\psi_r(x)$. Since it is easier to start with, we'll begin with calculating the auto-correlation function,
\begin{align}
\label{eqn:16}
{\cal C}_{\psi_r}(x-x';\alpha) =\; &k^2 \int \left< n(x+\alpha z) n(x'+\alpha z) \right > dz \nonumber \\
&- 2 k^2 \sum_{j=1}^N w_j \int \left < n(x+\alpha z) n(x' + \alpha_j z) \right > dz \nonumber \\ 
 &+ k^2  \sum_{j=1}^N \sum_{j'=1}^N w_j w_{j'} \int \left < n(x+\alpha_j z) n(x'+\alpha_{j'} z) \right > dz .
\end{align}
The variance is
\begin{align}
\label{eqn:variance}
{\cal C}_{\psi_r}(0;\alpha) = \;&k^2 \int \left< n(x)^2 \right > dz  \nonumber \\
&- 2 k^2 \sum_{j=1}^N w_j \int \left < n(\alpha z) n(\alpha_j z) \right > dz \nonumber \\ 
 &+ k^2  \sum_{j=1}^N \sum_{j'=1}^N  w_j w_{j'} \int \left < n(\alpha_j z) n(\alpha_{j'} z) \right > dz 
\end{align}
so, using (\ref{eqn:DCrelation}) the structure function is
\begin{align}
{\cal D}_{\psi_r}(x-x';\alpha) =\; &2\; {\cal C}_{\psi_r}(0;\alpha) - 2 \;{\cal C}_{\psi_r}(x-x';\alpha) \nonumber \\
= \;& k^2 \int \left < [n(x) - n(x')]^2 \right> dz \nonumber \\
&+ 2 \sum_{j=1}^N w_j \; k^2\int \left < \left [ n(\alpha z) - n(\alpha_{j} z) \right]^2 \right > \nonumber \\
&- 2  \sum_{j=1}^N w_j \; k^2\int \left < \left [ n(x+\alpha z) - n(x'+\alpha_{j} z) \right]^2 \right > \nonumber \\
&-   \sum_{j=1}^N \sum_{j'=1}^N w_j w_{j'} \;  k^2\int \left < \left [ n(\alpha_j z) - n(\alpha_{j'} z) \right]^2 \right > \nonumber \\
&+   \sum_{j=1}^N \sum_{j'=1}^N w_j w_{j'} \; k^2\int \left < \left [ n(x+\alpha_j z) - n(x'+\alpha_{j'} z) \right]^2 \right > .
\end{align}
Applying (\ref{eqn:SFindex}) we have
\begin{align}
\label{eqn:19}
{\cal D}_{\psi_r}(x-x';\alpha) =\; 2.914\; k^2 \int C_N^2(z) \bigg[ &\left | x - x' \right |^{5/3} \nonumber \\
&+ 2 \sum_{j=1}^N w_j \left | (\alpha - \alpha_j) z \right |^{5/3} \nonumber \\
&- 2 \sum_{j=1}^N w_j \left | x-x'+(\alpha - \alpha_j) z \right |^{5/3} \nonumber \\
&- \sum_{j=1}^N \sum_{j'=1}^N w_j w_{j'} \left | (\alpha_j - \alpha_{j'}) z \right |^{5/3}  \nonumber \\
&+ \sum_{j=1}^N \sum_{j'=1}^N w_j w_{j'} \left | x - x' + (\alpha_j - \alpha_{j'}) z \right |^{5/3} \bigg] dz
\end{align}

We now introduce some normalizations that will allow the above to be computed for a general case that can be scaled appropriately for specific cases. First of all consider the situation of zero guide stars, i.e. no correction. ${\cal D}_{\psi_r}$ reduces to ${\cal D}_\psi$, the structure function of the atmosphere and
\begin{equation}
\label{eqn:17}
{\cal D}_{\psi_r} (x-x';\alpha) = 2.914 \; k^2 \int C_N^2(z) \left | x-x' \right|^{5/3} dz = 6.88 \left( \left | x - x' \right| / r_0 \right)^{5/3}
\end{equation}
where
\begin{equation}
r_0 = \left [ 0.423 k^2 \int C_n^2(z) dz \right]^{-3/5}
\end{equation}
which is the familiar Fried seeing parameter. This suggests that we should change coordinates to units of $r_0$: $\mu = x / r_0$ then do the hard calculations in the normalized space so that particular cases are general results scaled appropriately by $r_0$.

There are three more normalizations that we will find useful. Let
\begin{equation}
\bar C_N^2(z) = C_N^2(z) \bigg/ \int C_N^2(z') dz'
\end{equation}
and
\begin{equation}
\label{eqn:etadef}
\eta = \alpha/\theta_0 \;;\quad \quad \eta_j = \alpha_j / \theta_0
\end{equation}
where $\theta_0$ is the isoplanatic angle 
\begin{equation}
\theta_0 = \left[ 2.914 k^2 \int C_N^2(z) z^{5/3} dz \right ] ^{-3/5}
\end{equation}
and
\begin{equation}
\label{eqn:xidef}
\xi = z / \bar h
\end{equation}
where $\bar h$ is the so-called mean height of turbulence:
\begin{equation}
\label{eqn:hbardef}
\bar h = \left [ \int \bar C_N^2 (z) z^{5/3} dz \right ]^{3/5}
\end{equation}
We will also take care to scale the function $C_N^2(z)$, which is a density in the altitude variable, for the new scaled altitude variable $\xi$
\begin{equation}
C_N^2(z) dz = C_N^2(\xi) d\xi \quad \text{or} \quad C_N^2(\xi) = C_N^2(z) \frac{dz}{d\xi} = C_N^2(z) \bar h
\end{equation}
which avoids having to carry around the extra factor of $\bar h$

We will use the fact that $\theta_0$ is proportional to $r_0$
\begin{equation}
\theta_0 = 0.314 \; r_0 / \bar h
\end{equation}
and also that
\begin{equation}
\int C_N^2 dz = \left [ 0.423 k^2 r_0^{5/3} \right ]^{-1}
\end{equation}

We now proceed with calculating the structure function in the normalized coordinates $\mu$, $\eta$. Equation (\ref{eqn:19}) in normalized form becomes
\begin{align}
\label{eqn:e27}
{\cal D}_{\psi_r}(\mu - \mu';\eta) &= \;6.88 \int \bar C_N^2 (\xi) \bigg[ (\mu - \mu')^{5/3} \nonumber \\
                       &- 2 \sum_{j=1}^N w_j \left(\left|\mu-\mu' + 0.314(\eta - \eta_j) \xi\right|^{5/3}
                       - \left |0.314(\eta - \eta_j) \xi\right|^{5/3}\right) \nonumber \\
                       &+ \sum_{j=1}^N \sum_{j'=1}^N w_j w_{j'} \left(\left|\mu-\mu' + 0.314\; (\eta_j - \eta_{j'})\xi\right|^{5/3}
                       - \left|0.314\; (\eta_j - \eta_{j'})\xi\right|^{5/3}\right)
                        \bigg] d \xi
\end{align}

For completeness, we also derive the normalized formula for the auto-correlation.  Starting with (\ref{eqn:16}) and using the fact that
$\sum_{j=1}^N w_j = 1$ implies ${\cal C}_N(0) = \sum_{j=1}^N w_j {\cal C}_n(0) = \sum_{j=1}^N \sum_{j'=1}^N w_j w_{j'} {\cal C}_n(0)$, we have:
\begin{align}
\label{eqn:variance_normalized}
{\cal C}_{\psi_r}(x-x';\alpha) =  \;& -k^2 \int \left[ {\cal C}_n(0) - {\cal C}_n(x-x') \right] dz \nonumber \\
&+ 2\,k^2 \sum_{j=1}^N w_j \int \left[ {\cal C}_n(0) - {\cal C}_n(x-x'+(\alpha-\alpha_j)z)\right] dz  \nonumber \\
&-k^2 \sum_{j=1}^N \sum_{j'=1}^N w_j w_{j'} \int \left[ {\cal C}_n(0) - {\cal C}_n(x-x'+(\alpha_j-\alpha_{j'})z)\right] dz  \nonumber \\
= \;& -\frac{1}{2} k^2 \int {\cal D}_n(x-x') dz \nonumber \\
& +k^2 \sum_{j=1}^N w_j \int {\cal D}_n(x-x'+(\alpha-\alpha_j)z) dz \nonumber \\
&-\frac{1}{2} k^2 \sum_{j=1}^N  \sum_{j'=1}^N w_j w_{j'} \int {\cal D}_n((\alpha_j-\alpha_j')z) \nonumber \\
=\;& 2.914 k^2 \int C_N^2(z) \bigg[ -\frac{1}{2} \left| x-x' \right|^{5/3}+\sum_{j=1}^N w_j \left| x-x'+(\alpha-\alpha_j) \right|^{5/3} \nonumber \\
&\qquad\qquad\qquad\qquad - \frac{1}{2} \sum_{j=1}^N \sum_{j'=1}^N w_j w_{j'} \left| x-x'+(\alpha_j - \alpha_j')z \right|^{5/3} \bigg] dz \nonumber \\
{\cal C}_{\psi_r}(\mu-\mu';\eta)=\;& 6.88 \int {\bar C}_N^2(\xi) \bigg [ -\frac{1}{2}\left| \mu-\mu'\right|^{5/3} +\sum_{j=1}^N w_j \left| \mu-\mu'+0.314(\eta-\eta_j)\xi \right|^{5/3} \nonumber \\
&\qquad\qquad\qquad\qquad - \frac{1}{2}  \sum_{j=1}^N \sum_{j'=1}^N w_j w_{j'} \left| \mu-\mu'+0.314(\eta_j - \eta_{j'})\xi \right|^{5/3} \bigg] dz
\end{align}

The argument $\mu-\mu'$ is 2-vector, spanning from 0 to spacings as large as the telescope diameter ($\left| \mu-\mu' \right| \leq D/r_0$). The arguments $\eta_i$ are 2-vectors denoting the positions of the guide stars, $i = 1,\ldots,N$. One can simplify the operation by assuming that the guide stars are in some symmetric geometry so that there would only be two parameters to describe the entire set: guide star spacing and number of guide stars, although a symmetric pattern that creates the best PSF is unclear at this point. We can think of a few simple geometries: 1) the guide stars are positioned on a circle around the center of the field, 2) the guide stars are positioned on a circle and there is one more guide star in the center, 3) the guide stars are positioned in a grid pattern out to the radius of the desired field. There are of course infinitely more possibilities.

Note that (\ref{eqn:e27}) and (\ref{eqn:variance_normalized}) can be written more compactly as
\begin{align}
{\cal D}_{\psi_r}(\mu - \mu'; \eta) = \;&6.88 \int \bar C_N^2 (\xi) \sum_{j=0}^N \sum_{j'=0}^N w_j w_{j'} \nonumber \\
&\times \left( \left |\mu-\mu' + 0.314\; (\eta_j - \eta_{j'})\xi\right|^{5/3} - \left |0.314\; (\eta_j - \eta_{j'})\xi\right|^{5/3} \right) d\xi
\end{align}
and
\begin{align}
{\cal C}_{\psi_r}(\mu-\mu';\eta) = \; -3.44 \int \bar C_N^2(\xi) \sum_{j=0}^N \sum_{j'=0}^N w_j w_j' 
\left| \mu-\mu'+0.314 \left( \eta_j - \eta_j' \right)\xi \right|^{5/3} d\xi
\end{align}
with $w_0 = -1$ and $\eta_0 = \eta$.

One can calculate these by first producing an $(N+1) \times (N+1)$ matrix of 2-d screens at each altitude $\xi$ and then summing (in a Reimann sum sense) over $\xi$. The screens are:
\begin{align}
&s_{jj'}(\mu,\eta,\xi) = 6.88 \left( \left |\mu + 0.314 (\eta_j-\eta_{j'})\xi\right|^{5/3} - \left |0.314 (\eta_j-\eta_{j'})\xi\right|^{5/3}\right) \\
&c_{{jj'}}(\mu, \eta,\xi) = -3.44 \left| \mu+0.314\left(\eta_j - \eta_{j'} \right)\xi \right|^{5/3} 
\end{align}
and then
\begin{align}
&{\cal D}_{\psi_r}(\mu;\eta) = \int \bar C_N^2(\xi) {\bf w}^T {\bf S}(\mu,\eta,\xi) {\bf w} d\xi  \label{eqn:quadratic_form_structure} \\
&{\cal C}_{\psi_r}(\mu;\eta) = \int \bar C_N^2(\xi) {\bf w}^T {\bf C}(\mu, \eta,\xi) {\bf w} d\xi \label{eqn:quadratic_form_correlation}
\end{align}
For notational compactness we will continue to use $\mu$ in place of $\mu-\mu'$ to indicate the separation of points in the horizontal plane. Within the matrices ${\bf S}$ and ${\bf C}$ only the first row and column change with the field position $\eta=\eta_0$. Furthermore, all the diagonal elements of the ${\bf S}$ are identical and equal to the open seeing structure function $6.88\; \mu^{5/3}$.

\subsection{Example calculations}

We consider two simple cases in order to spot check these formulas against previously known solutions. First we put all the turbulence at the ground level
\begin{equation}
\bar C_N^2(\xi) = \delta(\xi)
\end{equation}
in which case
\begin{equation}
{\cal D}_{\psi_r}(\mu;\eta) = 6.88 \sum_{j=0}^N \sum_{j'=0}^N w_j w_{j'} \left | \mu \right |^{5/3} = 6.88  \left | \mu \right |^{5/3} \sum_{j=0}^N w_j \sum_{j'=0}^N w_{j'}
\end{equation}
But
\begin{equation}
\sum_{j=0}^N w_j = w_0 + \sum_{j=1}^N w_j = -1 + 1 = 0
\end{equation}
where we have used the fact that the guide star weights must sum to one, and $w_0 = -1$. Therefore the structure function is zero, regardless of field direction or locations of the guide stars, as expected for when all the turbulence aberration is at the ground layer.

Next we consider a single guide star at position $\eta_1$ with respect to the field position at $\eta_0=0$. The quadratic forms (double-sum) inside the
structure function integral (\ref{eqn:quadratic_form_structure}) and the variance integral (\ref{eqn:quadratic_form_correlation}) (with $\mu=0$) are respectively
\begin{align} {\bf w}^T& {\bf S}(\mu,\eta,\xi) {\bf w} = \nonumber \\
(-1 \ 1) &\begin{pmatrix} \left| \mu \right|^{5/3} & \left| \mu - 0.314 \eta_1 \xi \right|^{5/3} - \left|0.314 \eta_1 \xi \right|^{5/3} \\
                                     \left| \mu + 0.314 \eta_1 \xi \right|^{5/3} - \left|0.314 \eta_1 \xi \right|^{5/3} & \left| \mu \right|^{5/3}     \end{pmatrix}
                                                                   \begin{pmatrix} -1 \\ 1 \end{pmatrix} \nonumber \\
            &= 2 \left| \mu \right|^{5/3} - \left| \mu - 0.314 \eta_1 \xi \right|^{5/3} + 2\left|0.314 \eta_1 \xi \right|^{5/3} - \left| \mu + 0.314 \eta_1 \xi \right|^{5/3}
\end{align}
and
\begin{align}
 {\bf w}^T {\bf C} (0,\eta,\xi) {\bf w} &= \nonumber \\
(-1 \ 1) &\begin{pmatrix} 0                                            & -\left| 0.314 \eta_1 \xi\right|^{5/3}      \\
                                     - \left| 0.314 \eta_1\xi \right|^{5/3} &                     0                              \end{pmatrix}
                                                                  \begin{pmatrix} -1 \\ 1 \end{pmatrix} \nonumber \\
 &= 6.88 \left| 0.314\, \eta_1 \xi \right|^{5/3}
\end{align}

We'll start with the variance. After integrating according to (\ref{eqn:quadratic_form_correlation}), the variance is:
\begin{align}
{\cal C}_{\psi_r}(0) &= \left( 3.44 \times 0.314^{5/3}\right)\, \eta_1^{5/3} \int \bar C_N^2(\xi) \left | \xi \right |^{5/3} d \xi \nonumber \\
&=  \eta_1^{5/3} = \left(\theta / \theta_0 \right) ^{5/3}
\end{align}
where we used the fact that definitions (\ref{eqn:xidef}) and (\ref{eqn:hbardef}) make the integral factor one, and then applied (\ref{eqn:etadef}) (with $\alpha_1 = \theta$) to write the solution in terms of $\theta_0$. This is the well known expression for the wavefront error variance due to angular anisoplanatism \cite{Hardy1998} (p103), as we would expect for correction with a single off-axis guide star.

The structure function is
\begin{align}
{\cal D}_{\psi_r}(u) = &6.88 \int \bar C_N^2(\xi) \nonumber \\
&\times \left [ 2 \left| \mu \right|^{5/3} - \left| \mu - 0.314 \eta_1 \xi \right|^{5/3} + 2\left|0.314 \eta_1 \xi \right|^{5/3} - \left| \mu + 0.314 \eta_1 \xi \right|^{5/3} \right ] d\xi
\end{align}
which, when substituting for the normalized parameters becomes
\begin{equation}
{\cal D}_{\psi_r}(r) = 2.914 \;k^2 \int C_n^2(z) \left [ 2 |r|^{5/3} + 2 \left | \theta z \right|^{5/3} - \left| r - \theta z \right |^{5/3} - \left| r + \theta z \right |^{5/3}  \right] dz.
\end{equation}
This is exactly the same as the structure function derived by Britton \cite{Britton2006} (equation 12) for the case of a single off-axis guide star.

\section{DM not conjugate to ground layer}
A common suggested implementation of GLAO is to have the telescope secondary be a deformable mirror.  The advantage is that no additional optics need to be inserted in the beam path from telescope to instrument to achieve AO correction. The adaptive secondary mirror (ASM) however is not conjugate to the ground so the applied wavefront correction will be at some optical distance that is either above or below the ground, depending on the telescope design. A Cassegrain telescope having a concave secondary intersecting the converging beam from the primary, has its secondary conjugate to below the ground. A Gregorian telescope having a convex secondary intersecting the primary beam after focus, has its secondary conjugate to above the ground.

To evaluate the behavior with an ASM, we need to dissect what we previously treated as ``GLAO wavefront correction'' into two component parts: the actual control as applied to the DM, and the computation of the control given the measured guide star wavefronts. We see that with the DM not at ground conjugate there is a shift of the beam on it depending on direction, but also we realize that in the control computation we have the freedom to shift wavefront measurements. Obviously once the telescope is built we can't change the conjugate height of the secondary, however we are free to change the control computations to, say, optimize performance under different $C_N^2$ profiles.

\subsection{Control at the DM}
First consider control applied to the adaptive secondary mirror. If the primary mirror is the entrance pupil, then beams from different angles are shifted on the ASM. Thus, for a science beam at direction $\alpha$
\begin{equation}
\label{eqn:ASMcontrol_priPupil}
\psi_r(x,\alpha) = \psi(x,\alpha) - \psi_d(x+\alpha z_c)
\end{equation}
where $z_c$ is the conjugate height of the secondary (positive for Gregorian, negative for Cassegrain). $\psi_d(x)$ is the wavefront correction applied to the ASM at physical position $x$ on the mirror surface.

In the case where the secondary mirror is the entrance pupil, then the beams counter-shift on the primary:
\begin{equation}
\label{eqn:ASMcontrol_secPupil}
\psi_r(x,\alpha) = \psi(x-\alpha z_c) - \psi_d(x)
\end{equation}
There is no difference in any of the wavefront cross-correlation statistics between these two entrance pupil cases. The affect is only a lateral ($x$) shift in the residual wavefront, but since the atmosphere statistics are assumed statistically stationary (statistics do not depend on the origin of coordinates), this does not affect the structure function. For the remainder, we proceed assuming the primary is the pupil.

\subsection{Shifting of wavefront measurements}
The second consideration is the controller. As an added degree of controller freedom, we allow shifts in the measured wavefronts before   averaging. The result is applied to the DM:
\begin{equation}
\label{eqn:shiftedWavefrontControl}
\psi_d(x) = \sum_{j=1}^N w_j \psi_j(x-\alpha_j z_s)
\end{equation}
The shift $\alpha_j z_s$ translates the measured wavefronts so that beams overlap at layer altitude $z_s$. If we average these wavefronts, we get a coherent addition of aberrations in layers near this layer altitude, within $\Delta z < r_0/(\alpha_j-\alpha_{j'})$, and an tendency to average out aberrations from layer altitudes outside this region. Of course, if $z_s=0$, the shift is zero, and the averaging occurs at the ground layer. This definition of $\Delta z$ can then be taken to be the thickness of the ground layer.

The motivation for an additional control degree of freedom is the possibility of improving AO correction over the field by emphasizing an altitude that is somewhat above ground, possibly nearer the dominant turbulent layer. Note that $z_s=0$ is the straight up averaging case, as we considered earlier. There is no particular reason at this point to set $z_l$ equal to the ASM conjugate height, but we may explore this possibility, especially in the Gregorian case where the conjugate height might be fortuitously close to the mean ground layer height.

\subsection{Non-ground-conjugate AO structure functions}
Substituting  (\ref{eqn:shiftedWavefrontControl}) into (\ref{eqn:ASMcontrol_priPupil}), and then recalling that $\psi(x) = k\int n(x+\alpha z,z) dz$ and $\psi_j(x) = k\int n(x+\alpha_j z,z) dz$,
\begin{align}
\psi_r(x,\alpha) &= \psi(x,\alpha) - \sum_{j=1}^N w_j \psi_j(x+\alpha z_c -\alpha_j z_s) \nonumber \\
&= k \int n(x+az,z) dz - k \sum_{j=1}^N w_j \int n(x+\alpha_j z+\alpha z_c-\alpha_j z_s,z) dz
\end{align}
Then we can calculate the correlation function
\begin{align}
{\cal C}_{\psi_r}(r;\alpha) = &k^2 \int {\cal C}_{n_z}(r) dz \nonumber \\
                                            &- k^2\sum_{j=1}^N w_j \int{\cal C}_{n_z}(r+(\alpha-\alpha_j)z-\alpha z_c+\alpha_j z_s)dz \nonumber \\
                                             & - k^2\sum_{j=1}^N w_j \int{\cal C}_{n_z}(r+(\alpha_j-\alpha)z+\alpha z_c-\alpha_j z_s)dz \nonumber \\
            & + k^2\sum_{j=1}^N\sum_{j'=1}^N w_j w_{j'} \int{\cal C}_{n_z}(r+(\alpha_j-\alpha_{j'})z -(\alpha_j-\alpha_{j'})z_s)dz
\end{align}
where $r = x-x'$ and ${\cal C}_{n_z}(r) = \left\langle n(x,z)n(x',z) \right\rangle$, and we have assumed there is no cross-correlation of index variation between layers at different altitudes $z$. With a bit of cleaver adding and subtracting within the arguments of the correlation functions, we can make the four terms look similar:
\begin{align}
{\cal C}_{\psi_r}(r;\alpha) = &k^2 \int {\cal C}_{n_z}(r+(\alpha-\alpha)(z-z_s)) dz \nonumber \\
                                            &- k^2\sum_{j=1}^N w_j \int{\cal C}_{n_z}(r+(\alpha-\alpha_j)(z-z_s) + \alpha (z_s - z_c))dz \nonumber \\
                                             & - k^2\sum_{j=1}^N w_j \int{\cal C}_{n_z}(r+(\alpha_j-\alpha)(z-z_s)-\alpha (z_s - z_c))dz \nonumber \\
            & + k^2\sum_{j=1}^N\sum_{j'=1}^N w_j w_{j'} \int{\cal C}_{n_z}(r+(\alpha_j-\alpha_{j'})(z-z_s))dz
\end{align}

In line with the thinking in appendix A.1, we can absorb the four terms into one double sum from $j=0$ to $N$ by defining $w_0 = -1$ and $\alpha_0 = \alpha$. We need to deal with the extra part in the cross-term's arguments and for this we utilize the Kroneker delta function.
\begin{equation}
{\cal C}_{\psi_r}(r;\alpha) = k^2 \sum_{j=0}^N \sum_{j'=0}^N 
                              w_j w_{j'}  \int  {\cal C}_{n_z}(r+(\alpha_j-\alpha_{j'})(z-z_s) + \alpha(z_s-z_c)(\delta_{j,0}-\delta_{j',0})) dz
\end{equation}
where $\delta_{j,0} = \{ 1 \text{ if } j=0; 0 \text{ otherwise}\}$. Finally, also following appendix A.1, we realize that
$\sum_{j=0}^N \sum_{j'=0}^N w_j w_{j'}  {\cal C}(0) = 0$ and ${\cal C}(r) = {\cal C}(0)-1/2\,{\cal D}(r)$, so we can express the correlation of $\psi_r$ in terms of structure functions of $n$
\begin{equation}
{\cal C}_{\psi_r}(r;\alpha) = -\frac{1}{2} k^2 \sum_{j=0}^N \sum_{j'=0}^N 
                              w_j w_{j'}  \int  {\cal D}_{n_z}(r+(\alpha_j-\alpha_{j'})(z-z_s) + \alpha(z_s-z_c)(\delta_{j,0}-\delta_{j',0})) dz
\end{equation}
where ${\cal D}_{n_z}(r) = \left\langle \left[ n(x,z)-n(x',z) \right]^2\right\rangle$

The structure function of the residual is
\begin{align}
{\cal D}_{\psi_r}(r;\alpha) = &2\,{\cal C}_{\psi_r}(0;\alpha) - 2\,{\cal C}_{\psi_r}(r;\alpha) \nonumber \\
        = k^2 \sum_{j=0}^N \sum_{j'=0}^N 
       &w_j w_{j'}  \int  \Big[ {\cal D}_{n_z}(r+(\alpha_j-\alpha_{j'})(z-z_s) + \alpha(z_s-z_c)(\delta_{j,0}-\delta_{j',0})) \nonumber \\
                            & - {\cal D}_{n_z}((\alpha_j-\alpha_{j'})(z-z_s) + \alpha(z_s-z_c)(\delta_{j,0}-\delta_{j',0})) \Big] dz                            
\end{align}
At this point we can substitute for the generic structure functions on the right hand side with the structure function for Kolmogorov index fluctuations
\begin{equation}
{\cal D}_{n_z}(r) = 2.914\, C_N^2(z) \left| r \right|^{5/3}
\end{equation}

We can reduce these equations to compact matrix forms:
\begin{align}
{\cal C}_{\psi_r}(r;\alpha) &= k^2\int C_N^2(z) {\bf w}^T {\bf C}(r;\alpha,z) {\bf w} dz \nonumber \\
{\cal D}_{\psi_r}(r;\alpha) &= k^2\int C_N^2(z) {\bf w}^T {\bf S}(r; \alpha,z) {\bf w} dz 
\end{align}
with matrix elements
\begin{align}
c_{jj'}(r,\alpha,z) = &-1.457\,\left | r+(\alpha_j-\alpha_{j'})(z-z_s) + \alpha(z_s-z_c)(\delta_{j,0}-\delta_{j',0}) \right |^{5/3} \nonumber \\
s_{jj'}(r,\alpha,z) = &\,2.914\,\Big(\left | r+(\alpha_j-\alpha_{j'})(z-z_s) + \alpha(z_s-z_c)(\delta_{j,0}-\delta_{j',0}) \right |^{5/3} \nonumber \\
              &- \left | (\alpha_j-\alpha_{j'})(z-z_s) + \alpha(z_s-z_c)(\delta_{j,0}-\delta_{j',0}) \right |^{5/3}\Big)
\end{align}

With the normalizations $\mu = r/r_0$, $\eta = \alpha/\theta_0$, $\xi = z/{\bar h}$, we have
\begin{align}
{\cal C}_{\psi_r}(\mu;\eta) &= \int {\bar C}_N^2(\xi) {\bf w}^T {\bar {\bf C}}(\mu;\eta,\xi) {\bf w} d\xi \nonumber \\
{\cal D}_{\psi_r}(\mu;\eta) &= \int {\bar C}_N^2(\xi) {\bf w}^T {\bar {\bf S}}(\mu;\eta,\xi) {\bf w} d\xi 
\end{align}
with matrix elements
\begin{align}
{\bar c}_{jj'}(\mu,\eta,\xi) = &-3.44 \left | \mu+0.314((\eta_j-\eta_{j'})(\xi-\xi_s) + \eta(\xi_s-\xi_c)(\delta_{j,0}-\delta_{j',0})) \right |^{5/3} \nonumber \\
{\bar s}_{jj'}(\mu,\eta,\xi) = &\;6.88 \,\Big( \left | \mu+0.314((\eta_j-\eta_{j'})(\xi-\xi_s) + \eta(\xi_s-\xi_c)(\delta_{j,0}-\delta_{j',0})) \right |^{5/3} \nonumber \\
               &-\left | 0.314( (\eta_j-\eta_{j'})(\xi-\xi_s) + \eta(\xi_s-\xi_c)(\delta_{j,0}-\delta_{j',0})) \right |^{5/3} \Big)
\end{align}

\section{Controller contribution to PSF}

Up to now, we have considered a perfect controller with the ability for instantaneous correction at all positions on the pupil. A real-world adaptive optics system will introduce errors here however, error due to a temporal delay in the application of wavefront control, and error in the wavefront fit by a deformable mirror that has only a finite number of actuators.

In both cases, there is an additional error introduced into the correction (\ref{eqn:residual_wf}):
\begin{equation}
\label{eqn:residual_wf2}
\psi_r(x) = \psi(x) - \sum_{j=1}^{N} w_j \psi_j(x) + \epsilon(x) = -\sum_{j=0}^N w_j \psi_j(x) + \epsilon(x) 
\end{equation}
where again we simplify the expression using $w_0 = -1$ and $\psi_0(x) = \psi(x)$. We first want to understand how this changes the PSF, based on how it changes the structure function. Using the addition rule (\ref{eqn:strucSum}) from appendix \ref{append:basic}
\begin{equation}
\label{eqn:e55}
{\cal D}_{\psi_r} (x-x') = {\cal D}_{\bar \psi_r} (x-x')  + 2{\cal D}_{\bar \psi_r \epsilon} (x-x') + {\cal D}_{\epsilon} (x-x')
\end{equation}
where $\bar \psi_r(x) = \psi_r(x)|_{\epsilon=0}$.
The effect is the introduction of two additional terms $2{\cal D}_{\bar \psi_r \epsilon} (x-x')$ and ${\cal D}_{\epsilon} (x-x')$ to the structure function. Since the MTF is the exponential of the structure function, the effect on the MTF is additional factors:
\begin{equation}
\label{eqn:MTFfactors}
MTF(u) = \tau_0(u) e^{-\frac{1}{2}{\cal D}_{\bar \psi_r}(u)}\;  e^{-{\cal D}_{\bar \psi_r \epsilon} (u)} \; e^{-\frac{1}{2}{\cal D}_{\epsilon} (u)}
\end{equation}
with each factor contributing to the degradation of the original unaberrated MTF, $\tau_0(u)$. Thus our task is to calculate the cross-term and the self-term for both the temporal and spatial error cases. 

\subsection{Temporal error}

Calculation and sampling delays in the control implementation introduce an error depending on the wind velocity

\begin{equation}
\epsilon(x) = k \sum_{j=1}^N w_j \int  \left( n(x + \alpha_j z) - n(x-v \Delta t + \alpha_j z)\right) dz
\end{equation}
where $v$ is the velocity of the wind (a 2-vector) and $\Delta t$ is the delay. From (\ref{eqn:SFderiv} in appendix \ref{append:deriv}, and recognizing that $\Delta x = - v \Delta t$,

\begin{align}
{\cal D}_{\psi_r}(r) = &{\cal D}_{\bar \psi_r}(r) + \big[ - 2 \sum_{j=1}^N w_j \nabla {\cal D}_{\psi\psi_j}(r) \nonumber \\
                                & \nabla{\cal D}_\psi (r) + \sum_{j=1}^N \sum_{j'=1}^N w_j w_{j'} \nabla {\cal D}_{\psi_j \psi_{j'}} (r) \big]\cdot (-v \Delta t)
\end{align}
When we define $w_0 = -1$, this simplifies to
\begin{equation}
{\cal D}_{\psi_r}(r) = {\cal D}_{\bar \psi_r}(r) - \sum_{j=0}^N \sum_{j'=0}^N w_j w_{j'} \nabla {\cal D}_{\psi_j \psi_{j'}} (r)\cdot v \Delta t
\end{equation}

\subsection{Spatial fitting error}

A deformable mirror has a finite number of actuators, in some sort of grid pattern, thus sampling the wavefront up to  an average spacing. To first order this acts as a spatial filter on the correction where only the spatial frequencies lower than one cycle per two spacings (the Nyquist limit) are addressed, and higher spatial frequencies remain mostly uncorrected. The correction error, $\epsilon(x)$ is

\begin{equation}
\epsilon(x) = \sum_{j=1}^N w_j \int h(x')  \psi_j(x-x') dx'
\end{equation}
where $h(x)$ is the high-pass filter that passes frequencies above Nyquist. From (\ref{eqn:55}) in appendix \ref{append:filter}:
\begin{align}
{\cal D}_{\psi_r}(r) = &{\cal D}_{\bar \psi_r}(r) 
   - 2\sum_{j=1}^N                        w_j          \int h(x')                \left[ {\cal D}_{\psi    \psi_j   }(x')      - {\cal D}_{\psi \psi_j      } (r-x')           \right ] dx' \nonumber \\
& + 2\sum_{j=1}^N \sum_{j'=1}^N w_j w_{j'} \int h(x')                \left[ {\cal D}_{\psi_j \psi_{j'}}(x')      - {\cal D}_{\psi_j \psi_{j'}} (r-x')            \right] dx' \nonumber \\
& - \sum_{j=1}^N \sum_{j'=1}^N w_j w_{j'} \int h(x') \int h(x'') \left[ {\cal D}_{\psi_j \psi_{j'}} (x''-x') - {\cal D}_{\psi_j \psi_{j'}} ( r + x'' - x') \right] dx' dx''
\end{align} 

\section{Sensitivity improvement calculation}

Intuitively, the sharpened PSF provides greater sensitivity for two reasons: 1) if is brighter near the center, providing greater signal against the background noise and 2) the total area covered by the PSF is smaller thus providing less total background noise in an aperture for photometry. The sensitivity is best quantified by a number that is proportional to the exposure time needed to achieve a measurement to a given signal-to-noise ratio, the ``equivalent noise area'' (ENA)\cite{King1983}
\begin{equation}
ENA = \frac{1}{\int PSF(\theta)^2 d\theta}
\end{equation}
In the case of background-limited observation (noise dominated by sky background), ENA is exactly inversely proportional to exposure time to a given SNR. It is also a reasonable approximation in the case of photon noise limited observation. ENA is a critical metric for comparing the performance of astronomical telescopes, instruments, and adaptive optics systems, as important as optical throughput and collecting aperture area\cite{Angeli2011}.

From Parseval's theorem, the ENA can be calculated from the MTF:
\begin{equation}
ENA = \frac{1}{\int MTF(u)^2 du}
\end{equation}
and we can thus use the MTF formulas like \ref{eqn:MTFfactors} to calculate ENA.

The speedup in exposure time provided by an AO system is the ratio of the ENAs:
\begin{equation}
{\rm Speedup} = \frac{ENA_{\rm seeing}}{ENA_{\rm AO}}
\end{equation}
 
 The ENA for open seeing is easily computed:
 \begin{align}
 ENA_{\rm seeing} &= \lambda^2 / \iint \exp\left\{-6.88 (|r|/r_0)^{5/3}\right\} d^2 r \nonumber \\
                               &= \frac{2^{16/5}}{\pi} \left( \frac{\lambda}{r_0}\right)^2 \nonumber \\
                               &\approx 2.92514\, \left(\frac{\lambda}{r_0} \right)^2
 \end{align}
 
 Since the ENA for GLAO is
 \begin{equation}
 ENA_{\rm AO} = \lambda^2 / \iint \exp\left\{-{\cal D}_{\psi_r}(r) \right\} d^2 r
 \end{equation}
 we can define a GLAO version of $r_0$ as
 \begin{equation}
 r_G = \frac{2^{8/5}}{\sqrt{\pi}} \sqrt{\iint \exp\left\{ -{\cal D}_{\psi_r}(r) \right\}d^2 r}
 \end{equation}
 
 Although the PSF for GLAO is not the same shape as that of open seeing with a scaled $r_0$, the effect on ENA is as if it were (with $r_G$ defined as above). The speedup factor is then
 \begin{equation}
 {\rm Seedup} = \left( \frac{r_G}{r_0} \right)^2
 \end{equation}

\section{Conclusion}

In this paper we have derived the analytic formulas for the (long exposure average) point spread function (PSF) for a ground-layer adaptive optics (GLAO) system. The PSF derives from the modulation transfer function (MTF) which in turn depends on second order statistics of the random wavefront phase variations, the phase structure function. To support these derivations, we developed algebraic and calculus rules for manipulating structure functions and their related entities, the cross-structure functions, which are summarized in the appendix. The PSF formulations account for the key aspects of the atmosphere and GLAO system: $C_n^2$ profile, guidestar constellation, DM conjugate altitude, DM actuator grid, and time delays in the controller. Adaptive optics correction improves science exposure sensitivity, that is it reduces the exposure time needed to achieve a given signal-to-noise ratio, by sharpening the PSF.  Sensitivity is inversely proportional to the equivalent noise area (ENA) of the PSF. GLAO's ENA can be calculated directly from either the MTF or PSF by squaring and integrating. The exposure time speedup from GLAO can be calculated by comparing its ENA to that of open seeing, which has a closed form equation. This lends itself naturally to the definition of an $r_0$-like quantity that characterizes the GLAO point spread function.


\bibliography{paper}

\pagebreak
\begin{appendix}
\section{Structure function calculus}
This appendix gives mathematical background for some of the derivations used in the main text.
\subsection{Basic arithmetic}
\label{append:basic}

{\it Structure and correlation functions: definitions and similarity properties}

The structure function is defined on a stationary process $a(x)$
\begin{align}
\label{eqn:structFcnDefn}
{\cal D}_a(r) &= \left< \left [ a(x) - a(x') \right ]^2 \right>  \nonumber \\
                            &= 2\;\left< a(x)^2 \right> - 2\; \left<a(x) a(x')\right> = 2\;{\cal C}_a (0) - 2\;{\cal C}_a (r)
\end{align}
where ${\cal C}_a (r) = \left < a(x) a(x') \right >$ is the auto-correlation function and $r = x-x'$. Stationarity implies that the auto-correlation (and structure function) is only a function of the difference vector of the two points in space, $x-x'$, and not on the absolute position in space.

The cross-structure function between two stationary processes $a(x)$ and $b(x)$ is defined
\begin{align}
{\cal D}_{a b}(r) &= \left< \left [ a(x) - a(x') \right ]\left [ b(x) - b(x')\right] \right>  \nonumber\\
                                             &= 2 \; \left<a(x) b(x)\right> - \left< a(x) b(x') \right> - \left< a(x') b(x) \right> \nonumber \\
                                             &= 2\;{\cal C}_{a b}(0) - {\cal C}_{a b }(r) - {\cal C}_{a b}(-r)
\end{align}
where ${\cal C}_{a b}(r) = \left< a(x) b(x')\right>$ is the cross-correlation function. The cross-correlation is symmetric under change of argument sign {\it and} indices
\begin{equation}
{\cal C}_{a b}(r) = {\cal C}_{ b a}(-r).
\end{equation}
while the cross-structure function is symmetric under change of sign {\it or} indices
\begin{equation}
{\cal D}_{a b}(r) = {\cal D}_{a b}(-r) = {\cal D}_{b a} (r) = {\cal D}_{b a} (-r)
\end{equation}
If the processes $a(x)$ and $b(x)$ are statistically independent and zero-mean, then their cross-correlation is zero, and so also is the cross-structure function.

\vspace{12pt}
\noindent{\it Structure functions of sums}

The structure function of a sum of dependent processes $c(x) = a(x) + b(x)$ is
\begin{align}
\label{eqn:strucSum}
{\cal D}_c (r) &= \left< \left [ a(x) + b(x) - a(x') - b(x') \right]^2 \right> \nonumber \\
                         &= \left< \left [ (a(x) - a(x')) + (b(x)-b(x')) \right]^2 \right> \nonumber \\
                         &= {\cal D}_a (r) + {\cal D}_b (r) + 2\; {\cal D}_{a b} (r).
\end{align}
If the processes $a(x)$ and $b(x)$ are statistically independent and zero-mean, then the structure function of the sum is the sum of the structure functions.

Generalizing to a sum of many processes:
\begin{equation}
c(x) = \sum_j a_j(x)
\end{equation}

\begin{equation}
\label{eqn:manySumRule}
{\cal D}_c(r) = \sum_j \sum_{j'} {\cal D}_{a_j a_{j'}} (r)
\end{equation}
with the convention that when $j=j'$, ${\cal D}_{a_j a_j}(r) = {\cal D}_{a_j}(r)$.

Next we consider a summation form that we encounter often in the main section of this report:

\begin{equation}
\label{eqn:A1_residualSum}
c(x) = a(x) - \sum_{j=1}^N w_j b_i(x)
\end{equation}
We make the following two assumptions. First, assume the weights, $w_j$ total to one:
\begin{equation}
\text{\bf  1)}\hspace{2.5cm} \sum_{j=1}^N w_j = 1  \hspace{3.5cm}
\end {equation}
Second, assume that the processes $a$ and $b_i$ are cross-correlated to each other through shift relations:
\begin{align}
\text{\bf  2)}\hspace{.1cm} \hspace{2.0cm} {\cal C}_{a b_j} &= {\cal C}_a(r+\xi_j) \hspace{3.0cm} \nonumber \\
{\cal C}_{b_j b_{j'}}(r) &= {\cal C}_a(r+\xi_j - \xi_{j'}) 
\end{align}

Using assumption 2, the auto-correlation function of $c(x)$ is
\begin{equation}
\label{eqn:A1_corResidual}
{\cal C}_c(r) = {\cal C}_a(r) - \sum_{j=1}^N w_j {\cal C}_a(r+\xi_j) - \sum_{j=1}^N w_j {\cal C}_a(r-\xi_j) + \sum_{j=1}^N \sum_{j'=1}^N w_j w_{j'} {\cal C}_a(r+\xi_j - \xi_{j'})
\end{equation}
where we used assumption 2 to substitute for all the cross-correlation terms. Next we apply assumption 1 and the fact that ${\cal C}_a(0)$ is a constant to realize that
\begin{equation}
\label{eqn:A1_corZero}
0 = {\cal C}_a(0) - \sum_{j=1}^N w_j {\cal C}_a(0) - \sum_{j=1}^N w_j {\cal C}_a(0) + \sum_{j=1}^N \sum_{j'=1}^N w_j w_{j'} {\cal C}_a(0)
\end{equation}
Subtracting (\ref{eqn:A1_corZero}) from (\ref{eqn:A1_corResidual}) and using the definition for structure function (\ref{eqn:structFcnDefn}) we have
\begin{equation}
\label{eqn:A1_finalCorResidual}
{\cal C}_c(r) = -\frac{1}{2} \left[ {\cal D}_a(r) - \sum_{j=1}^N w_j {\cal D}_a(r+\xi_j) - \sum_{j=1}^N w_j {\cal D}_a(r-\xi_j) + \sum_{j=1}^N \sum_{j'=1}^N w_j w_{j'} {\cal D}_a(r+\xi_j - \xi_{j'})\right ]
\end{equation}
which is the correlation function of the residual in terms of the structure functions of the components. Note that the assumption that the weights sum to one means that there are no $a$ variance terms (${\cal C}_a(0)$) in the result. Thus the formula is valid for any, even arbitrarily large, initial $a$ variance, as is often the case with a fractal process such as Kolmogorov turbulence. The resulting process $c$ has  a finite variance
\begin{equation}
\label{eqn:A1_finalVariance}
\sigma_c^2 = {\cal C}_c(0) = \frac{1}{2} \left[ \sum_{j=1}^N w_j {\cal D}_a(\xi_j) + \sum_{j=1}^N w_j {\cal D}_a(-\xi_j) - \sum_{j=1}^N \sum_{j'=1}^N w_j w_{j'} {\cal D}_a(\xi_j - \xi_{j'})\right ]
\end{equation}

We can now use (\ref{eqn:structFcnDefn}) to form the expression for the structure function of $c(x)$:
\begin{align}
{\cal D}_c(r) = &2\,{\cal C}_a(0) - 2\,{\cal C}_a(r) \nonumber \\
= &{\cal D}_a(r) 
  - \sum_{j=1}^N w_j ({\cal D}_a(r+\xi_j)-{\cal D}_a(\xi_j)) 
  - \sum_{j=1}^N w_j ({\cal D}_a(r-\xi_j)-{\cal D}_a(\xi_j)) \nonumber \\
&+\sum_{j=1}^N \sum_{j'=1}^N w_j w_{j'} ({\cal D}_a(r+\xi_j - \xi_{j'})-{\cal D}_a(\xi_j - \xi_{j'})) \label{eqn:A1_finalStrucFcn}
\end{align}

Equations (\ref{eqn:A1_finalCorResidual}) and (\ref{eqn:A1_finalStrucFcn}) can be further simplified if we define $w_0 = -1$ and $\xi_0 = 0$. Then all the terms can be combined into one double sum from $j=0$ to $N$:
\begin{align}
{\cal C}_c(r) &= -\frac{1}{2} \sum_{j=0}^N \sum_{j'=0}^N w_j w_{j'} {\cal D}_a(r+\xi_j - \xi_{j'}) \\
{\cal D}_c(r) &= \sum_{j=0}^N \sum_{j'=0}^N w_j w_{j'} ({\cal D}_a(r+\xi_j - \xi_{j'})-{\cal D}_a(\xi_j - \xi_{j'})) 
\end{align}
which can be written compactly as matrix quadratic forms
\begin{align}
{\cal C}_c(r) &= -\frac{1}{2} {\bf w}^T  {\bf C} {\bf w} \\
{\cal D}_c(r) &= {\bf w}^T {\bf S} {\bf w}
\end{align}
where
\begin{align}
{\bf C}_{jj'} &= {\cal D}_a(r+\xi_j - \xi_{j'}) \\
{\bf S}_{jj'} &= {\cal D}_a(r+\xi_j - \xi_{j'})-{\cal D}_a(\xi_j - \xi_{j'})
\end{align}
\subsection{Derivatives}
\label{append:deriv}
We evaluate the cross-structure function of a process $a(x)$ with its derivative $\nabla a(x)$:
\begin{equation}
\label{eqn:crossStructDeriv}
{\cal D}_{a \;\nabla a}(r) = \left< \left [ a(x) - a(x') ] [ \nabla a(x) - \nabla a(x') \right] \right>
\end{equation}

Start by considering the derivative of the structure function itself, with respect to its argument $r$. This can be expressed as the derivative at the end point $x$
\begin{equation}
\nabla_r{\cal D}_a(r) = \nabla_x {\cal D}(x-x') = 2\left<\left[ a(x)-a(x') \right]\nabla_x a(x)\right>
\end{equation}
But by symmetry, since ${\cal D}$ is only a function of the difference $x-x'$ we can also express it as its derivative at the other end point $x'$
\begin{equation}
\nabla_r{\cal D}_a(r) = -\nabla_{x'} {\cal D}(x-x') = -2\left<\left[ a(x)-a(x') \right]\nabla_{x'} a(x')\right>
\end{equation}
Since these two are equal, we can take their average
\begin{equation}
\nabla_r{\cal D}_a(r) = \left<\left[a(x)-a(x')\right]\left[\nabla_x a(x) - \nabla_{x'} a(x')\right]\right>
\end{equation}
Comparing to \ref{eqn:crossStructDeriv}, we have therefore
\begin{equation}
\label{eqn:derivSF}
{\cal D}_{a \;\nabla a}(r) = \nabla {\cal D}_a(r)
\end{equation}

Using similar manipulations, the derivative of the cross-structure function of $a(x)$ and $b(x)$ is
\begin{equation}
\label{eqn:derivCross}
\nabla {\cal D}_{ab}(r) = \frac{1}{2} {\cal D}_{a \; \nabla b}(r) + \frac{1}{2} {\cal D}_{\nabla a \; b}(r).
\end{equation}

There is a case considered in the main text where the second term in the difference $a(x)-b(x)$ undergoes a first-order perturbation
\begin{align}
\label{eqn:expansion1}
c(x) &= a(x) - b(x+\Delta x) \nonumber \\
       &= a(x) - b(x) - \nabla b(x) \Delta x.
\end{align}
Because the processes $a(x)$ and $b(x)$ are stationary, the statistics of $c(x)$ and $c(x+q)$ are the same regardless of shift $q$. So we shift the entire relation (\ref{eqn:expansion1}) in space by $-\Delta x$, then expand on $a(x)$ instead of $b(x)$:
\begin{align}
\label{eqn:expansion2}
c(x-\Delta x) &= a(x-\Delta x) - b(x) \nonumber \\
                     &= a(x) - \nabla a(x) \Delta x - b(x).
\end{align}

This bit of symmetry allows us to write a structure function for the process $c(x)$ in terms of the structure functions, cross-structure functions, and their derivatives. This will be shown below.

Applying (\ref{eqn:strucSum}) to (\ref{eqn:expansion1}) and dropping the term that is second order in $\Delta x$,
\begin{align}
{\cal D}_c (r) = &{\cal D}_a (r) +  {\cal D}_b(r) - 2\; {\cal D}_{a b} (r) \nonumber \\
                     & - 2 \; {\cal D}_{a\;\nabla b}(r) \Delta x + 2 \; {\cal D}_{b \;\nabla b}(r) \Delta x.
\end{align}
Similarly, applying (\ref{eqn:strucSum})  to (\ref{eqn:expansion2}) and dropping the second order term.
\begin{align}
{\cal D}_c (r) = &{\cal D}_a (r) + {\cal D}_b(r) - 2\; {\cal D}_{a b}(r) \nonumber \\
                     & - 2 \; {\cal D}_{\nabla a \; b}(r) \Delta x + 2\; {\cal D}_{a \; \nabla a}(r) \Delta x.
\end{align}
Since these two quantities are equal, we can take their average:
\begin{align}
{\cal D}_c (r) =& {\cal D}_a (r) + {\cal D}_b(r) - 2\; {\cal D}_{a b}(r) \nonumber \\
                     & + \left[ - {\cal D}_{a \; \nabla b}(r) - {\cal D}_{\nabla a\; b}(r) + {\cal D}_{a \; \nabla a}(r) + {\cal D}_{b \; \nabla b}(r) \right] \Delta x.
\end{align}
Then, substituting (\ref{eqn:derivSF}) and (\ref{eqn:derivCross}), this becomes
\begin{align}
\label{eqn:SFderiv}
{\cal D}_c (r) =& {\cal D}_a (r) + {\cal D}_b(r) - 2\; {\cal D}_{a b}(r) \nonumber \\
                       & + \left[ -2\nabla {\cal D}_{ab}(r) + \nabla {\cal D}_{a}(r) + \nabla {\cal D}_{b}(r) \right] \Delta x
\end{align}
which gives the structure function of the perturbed sum in terms of the original structure functions and their derivatives.

In the special case where the structure functions are of the form
\begin{equation}
{\cal D}(r) = \left| r + \gamma \right|^{5/3} + \left| \gamma \right|^{5/3}
\end{equation}
the ordinary derivative exists at every point except ${\bf r}=-{\boldsymbol \gamma}$:
\begin{equation}
\nabla {\cal D}(r) = \frac{5}{3} \left| r + \gamma \right|^{2/3} \frac{\bf r+\boldsymbol\gamma}{|r+\gamma|}
=\frac{5}{3} \left| r + \gamma \right|^{-1/3} \left({\bf r}+{\boldsymbol\gamma}\right)
\end{equation}
where the bold font indicates that the gradient is of course a vector quantity. What do we do about the bad point ${\bf r}=-{\boldsymbol \gamma}$? The derivative is actually well-behaved there as its magnitude decreases uniformly to zero as ${\bf r}\rightarrow -{\boldsymbol\gamma}$ from any direction. Therefore we set
\begin{equation}
\nabla {\cal D}(r)|_{\bf r=-\boldsymbol \gamma} = 0
\end{equation} 
\subsection{Filtering}
\label{append:filter}

Another situation that is encountered is when the second term in a difference is spatial-filtered
\begin{equation}
c(x) = a(x) - f(x) \ast b(x) = a(x) - \int f(x-x'') b(x'') d x'',
\end{equation}
keeping in mind that this is a two-dimensional integral when $x$ is a 2-vector.

First, we'll rewrite this into a form that will allow convenient separation of terms later
\begin{equation}
c(x) = a(x) - b(x) + \int h(x'') b(x-x'') d x''
\end{equation}
where $h(x) = \delta(x) - f(x)$. We will require that $\left|f(x)\right| \rightarrow 0$ (and therefore $\left|h(x)\right| \rightarrow 0$) faster than $\left|x\right|^{-2/3}$ as $\left|x\right| \rightarrow \infty$. This is true in practical cases where for, example, the response falls as a Gaussian, or as $\sin(x)/x$.

Next, we write the auto correlation function
\begin{align}
{\cal C}_c (r) =&\; {\cal C}_a (r) + {\cal C}_b (r) - {\cal C}_{a b}(r) - {\cal C}_{b a}(r) \nonumber \\
                       &+ \int h(x'') \left[\left< a(x) b(x'-x'')\right> - \left< b(x) b(x'-x''')\right> \right]dx'' \nonumber \\
                       &+ \int h(x'') \left[\left< a(x') b(x-x'')\right> - \left< b(x') b(x-x''')\right> \right]dx'' \nonumber \\
                       &+ \int h(x'') \int h(x''') \left< b(x-x'') b(x'-x''')\right> dx'' dx''' \nonumber \\
                      =&\;{\cal C}_a (r) + {\cal C}_b (r) - {\cal C}_{a b}(r) - {\cal C}_{ba}(r) \nonumber \\
                       &+\int h(x'') \left[ {\cal C}_{ab} (r+x'') - {\cal C}_b(r+x'') \right] dx'' \nonumber \\
                       &+\int h(x'') \left[ {\cal C}_{ab} (-r+x'') - {\cal C}_b(-r+x'') \right] dx'' \nonumber \\
                       &+\int h(x'') \int h(x''') {\cal C}_b(r-x''+x''') dx'' dx'''
\end{align}
and the variance, which is the auto correlation at $r=0$
\begin{align}
{\cal C}_c(0) = &\; {\cal C}_a(0) + {\cal C}_b (0) - 2\; {\cal C}_{a b}(0) \nonumber \\
                       &+2\int h(x'') \left[ {\cal C}_{ab} (x'') - {\cal C}_b(x'') \right] dx'' \nonumber \\
                       &+\int h(x'') \int h(x''') {\cal C}_b(-x''+x''') dx'' dx'''.
\end{align}
The resulting structure function is
\begin{align}
\label{eqn:67}
{\cal D}_c(r) =& \; 2\; {\cal C}_c (0) - 2\; {\cal C}_c(r) \nonumber \\
                     =&\; {\cal D}_a(r) + {\cal D}_b (r) - 2\; {\cal D}_{ab} (r) \nonumber \\
                       & + 2 \int h(x'') \left[ {\cal C}_{ab}(x'') - {\cal C}_{ab}(r+x'')\right] dx'' \nonumber \\
                       & + 2 \int h(x'') \left[ {\cal C}_{ab}(x'') - {\cal C}_{ab}(-r+x'')\right] dx'' \nonumber \\
                       & - 2 \int h(x'') \left[  {\cal C}_{b}(x'') - {\cal C}_b(-r+x'')\right] dx'' \nonumber \\
                       & - 2 \int h(x'') \left[  {\cal C}_{b}(x'') - {\cal C}_b(-r+x'')\right] dx'' \nonumber \\
                       & + 2 \int h(x'') \int h(x''') \left[{\cal C}_b(-x''+x''') - {\cal C}_b(r-x''+x''')\right] dx'' dx'''.
\end{align}
Now adding and subtracting constant terms ${\cal C}_b(0)$ and ${\cal C}_{ab}(0)$ under the integrals which, in toto, cancel out:
\begin{align}
{\cal D}_c(r) =&\; {\cal D}_a(r) + {\cal D}_b (r) - 2\; {\cal D}_{a b} (r) \nonumber \\
                       & + 2 \int h(x'') \left[ ({\cal C}_{ab}(x'') - {\cal C}_{ab}(0)) - ({\cal C}_{ab}(r+x'') - {\cal C}_{ab}(0))\right] dx'' \nonumber \\
                       & + 2 \int h(x'') \left[ ({\cal C}_{ab}(x'') - {\cal C}_{ab}(0)) - ({\cal C}_{ab}(-r+x'') - {\cal C}_{ab}(0))\right] dx'' \nonumber \\
                       & - 2 \int h(x'') \left[ ({\cal C}_{b}(x'') -{\cal C}_b(0)) - ({\cal C}_b(-r+x'') - {\cal C}_b(0))\right] dx'' \nonumber \\
                       & - 2 \int h(x'') \left[ ({\cal C}_{b}(x'') -{\cal C}_b(0)) - ({\cal C}_b(-r+x'') - {\cal C}_b(0))\right] dx'' \nonumber \\
                       & + 2 \int h(x'') \int h(x''') \left[({\cal C}_b(-x''+x''') - {\cal C}_b(0)) - ({\cal C}_b(r-x''+x''')-{\cal C}_b(0))\right] dx'' dx'''.
\end{align}
Finally, we have
\begin{align}
\label{eqn:55}
{\cal D}_c(r) =&\; {\cal D}_a(r) + {\cal D}_b (r) - 2\; {\cal D}_{ab} (r) \nonumber \\
                       & - 2\int h(x'') \left[ {\cal D}_{ab}(x'') - {\cal D}_{ab}(r-x'') \right] d x''
                        + 2\int h(x'') \left[ {\cal D}_{b}(x'') - {\cal D}_{b}(r-x'') \right] d x'' \nonumber \\
                       & - \int h(x'') \int h(x''') \left[ {\cal D}_b (x''-x''') - {\cal D}_b (r - x'' + x''') \right] dx'' dx'''.
\end{align}
Note that in each integrand, the structure function factor, a difference of structure functions offset by $r$, tends to an asymptote of $(d{\cal D}/d x) \times r$ as the independent variable ($x''$ or $x'''$) goes to magnitude infinity. Thus the response function $h(x)$ is required to go to zero at a rate faster than $(d{\cal D}/d x)^{-1}$ in order for the integrals to converge.

\subsection{Exact expressions for the funny constants\cite{Hardy1998}}
\label{append:funny}

\begin{align}
``2.914"&= \frac{1}{3}\;\frac{6}{5}\;
\frac {\Gamma(\left(\frac{1}{2}\right) \Gamma\left(\frac{1}{6}\right)} {\Gamma\left(\frac{2}{3}\right) }
= 2.914380777465098  \quad {\rm (Hardy, p.90)}\\
&\quad \nonumber \\
``6.88" &= 8 \sqrt{2} \left[ \frac{3}{5}\Gamma\left(\frac{6}{5}\right)\right ]^{5/6} = 6.883877182293811 \quad {\rm (Hardy, p.91)} \\
&\quad \nonumber \\
``0.314" &= \left (8 \sqrt{2} \left[ \frac{3}{5}\Gamma\left(\frac{6}{5}\right)\right ]^{5/6}\right)^{-3/5} = 0.3142679785118208 \quad {\rm (Hardy, p. 103)}
\end{align}

\end{appendix}

\end{document}